\documentclass[11pt, a4paper, Physsubmission, Phys]{SciPost} 

\hypersetup{
    colorlinks,
    linkcolor={red!50!black},
    citecolor={blue!50!black},
    urlcolor={blue!80!black}
}

\usepackage[bitstream-charter]{mathdesign}
\usepackage{here} 
\DeclareSymbolFont{usualmathcal}{OMS}{cmsy}{m}{n}
\DeclareSymbolFontAlphabet{\mathcal}{usualmathcal}

\begin{document}

\begin{center}{\Large \textbf{
The ALPACA experiment: The project of the first sub-PeV gamma-ray observation in the southern sky
}}\end{center}

\begin{center}
Teruyoshi Kawashima\textsuperscript{1$\star$} for The ALPACA Collaboration
\end{center}

\begin{center}
{\bf 1} Institute for Cosmic Ray Research, the University of Tokyo, 5-1-5, Kashiwanoha, Kashiwa-shi, Chiba, 277-8582, Japan 
\\
* tkawa4ma@icrr.u-tokyo.ac.jp
\end{center}

\begin{center}
\today
\end{center}


\definecolor{palegray}{gray}{0.95}
\begin{center}
\colorbox{palegray}{
  \begin{tabular}{rr}
  \begin{minipage}{0.1\textwidth}
    \includegraphics[width=30mm]{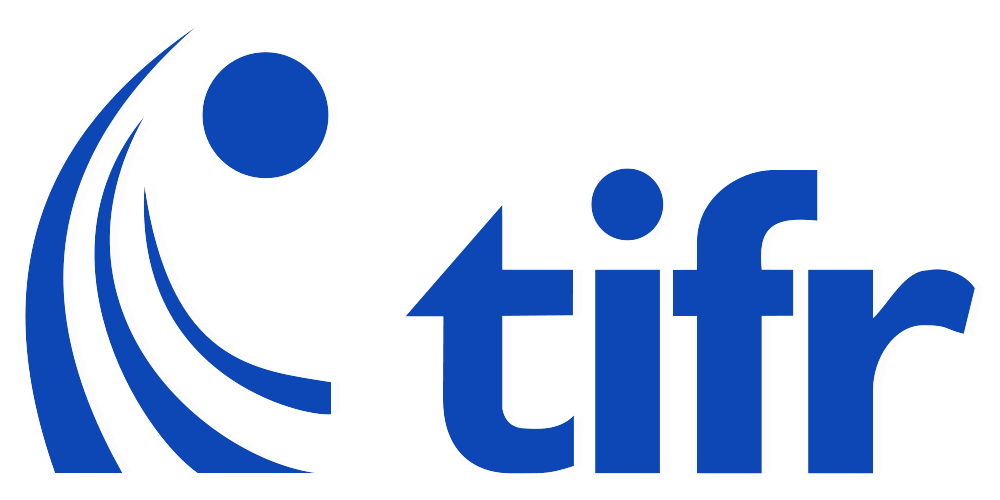}
  \end{minipage}
  &
  \begin{minipage}{0.85\textwidth}
    \begin{center}
    {\it 21st International Symposium on Very High Energy Cosmic Ray Interactions (ISVHE- CRI 2022)}\\
    {\it Online, 23-28 May 2022} \\
    \doi{10.21468/SciPostPhysProc.?}\\
    \end{center}
  \end{minipage}
\end{tabular}
}
\end{center}

\section*{Abstract}
{\bf
The ALPACA experiment is a project aiming to observe sub-PeV gamma rays for the first time in the southern hemisphere. The main goal of ALPACA is to identify PeVatrons, the accelerators of Galactic PeV cosmic rays, by observing sub-PeV pion-decay gamma rays generated in interactions between PeV cosmic rays and the interstellar medium. This new air shower experiment is located at an altitude of 4,740\,m above sea level in the middle of Mt. Chakartaya in Bolivia. The air shower array consists of 401 scintillation counters covering an 83,000\,m$^2$ surface area. In addition, a water-Cherenkov-type muon detector array with an area of 3,700\,m$^2$ is installed to discriminate gamma rays from background cosmic rays. The prototype array ALPAQUITA will start data taking in 2022 and will extend to ALPACA in 2024. We report on a general introduction to ALPACA, the progress of the project, and the sensitivity to sub-PeV gamma rays.
}

\vspace{10pt}
\noindent\rule{\textwidth}{1pt}
\tableofcontents\thispagestyle{fancy}
\noindent\rule{\textwidth}{1pt}
\vspace{10pt}

\section{Introduction}
\label{sec:intro}
The identity of the PeV-cosmic-ray accelerators in our galaxy, called PeVatrons, is unknown, and identifying them is an important challenge in astrophysics. When PeV cosmic rays interact with the interstellar medium, neutral pions are produced, which immediately decay into two photons above 100 TeV (sub-PeV gamma rays). Therefore, observations of sub-PeV gamma rays are essential to the search for PeVatrons and have been actively performed in recent years. 
The Tibet AS$\gamma$ collaboration opened the door to sub-PeV gamma-ray astronomy for the first time with the detection of sub-PeV gamma rays from the Crab Nebula\cite{tibet_crab} followed by HAWC\cite{hawc_crab} and LHAASO\cite{lhaaso_crab}.
Then, sub-PeV gamma-ray observations from several sources were reported by these collaborators, such as SNR G106.3+2.7, Cygnus Cocoon, and J1908+0621\cite{tibet_g106,hawc_cygnus,lhaaso_j1908}. In addition, the detection of diffuse sub-PeV gamma rays by the Tibet AS$\gamma$ collaboration has established the existence of PeVatrons in our galaxy\cite{tibet_diffuse}. Thus, the sub-PeV gamma-ray astronomy in the Northern Hemisphere has been greatly developed by these collaborators. However, the direct identification of PeVatrons is still to be achieved.
 
As reported in the H.E.S.S. Galactic plane survey\cite{hess_survey}, more high-energy gamma-ray sources exists in the southern sky including some promising PeVatron candidates. The H.E.S.S. group reported gamma-ray emissions of up to 40 TeV (spectral index:~$-2.3$) from the Galactic Center region, strongly pointing out the possibility of the Galactic Center being a PeVatron\cite{hess_gc}. Furthermore, the H.E.S.S. group reported the existence of an unidentified object HESS J1702-420A, a PeVatron candidate having a hard gamma-ray spectrum (spectral index:~$-1.5$) extending above 100\,TeV without cutoff\cite{hess_j1702}. However, in the southern sky, sub-PeV gamma-ray observations are still almost untouched, so it is important to start surveys.

The Andes Large area PArticle detector for Cosmic ray physics and Astronomy (ALPACA) collaboration aims to demonstrate the first sub-PeV gamma ray observations in the southern hemisphere using an air-shoer-array detector. The array is going to be construct in Bolivia and plans to start observations in 2024. The construction of ALPAQUITA, a prototype array of ALPACA, is currently underway and is scheduled to start observations in 2022. This paper presents a general introduction to the ALPACA project, the sensitivites of ALPACA and ALPAQUITA to celestial gamma rays, and the construction status of ALPAQUITA.

\section{Outline of the ALPACA project}
\label{sec:outline}
The ALPACA project is an international collaboration between Bolivia, Mexico, and Japan. A new array based on Tibet AS$\gamma$ design\cite{tibet_tsako} is constructed at an altitude of 4,740\,m above sea level in the middle of Mt. Chacaltaya, Bolivia (16$^\circ$23'\,S,~68$^\circ$08'\,W and the atmospheric depth of $572.4\,\mathrm{g/cm^2}$).
\subsection{ALPACA}
\label{sec:alpaca}
\begin{figure}[h]
\centering
\includegraphics[width=0.65\textwidth]{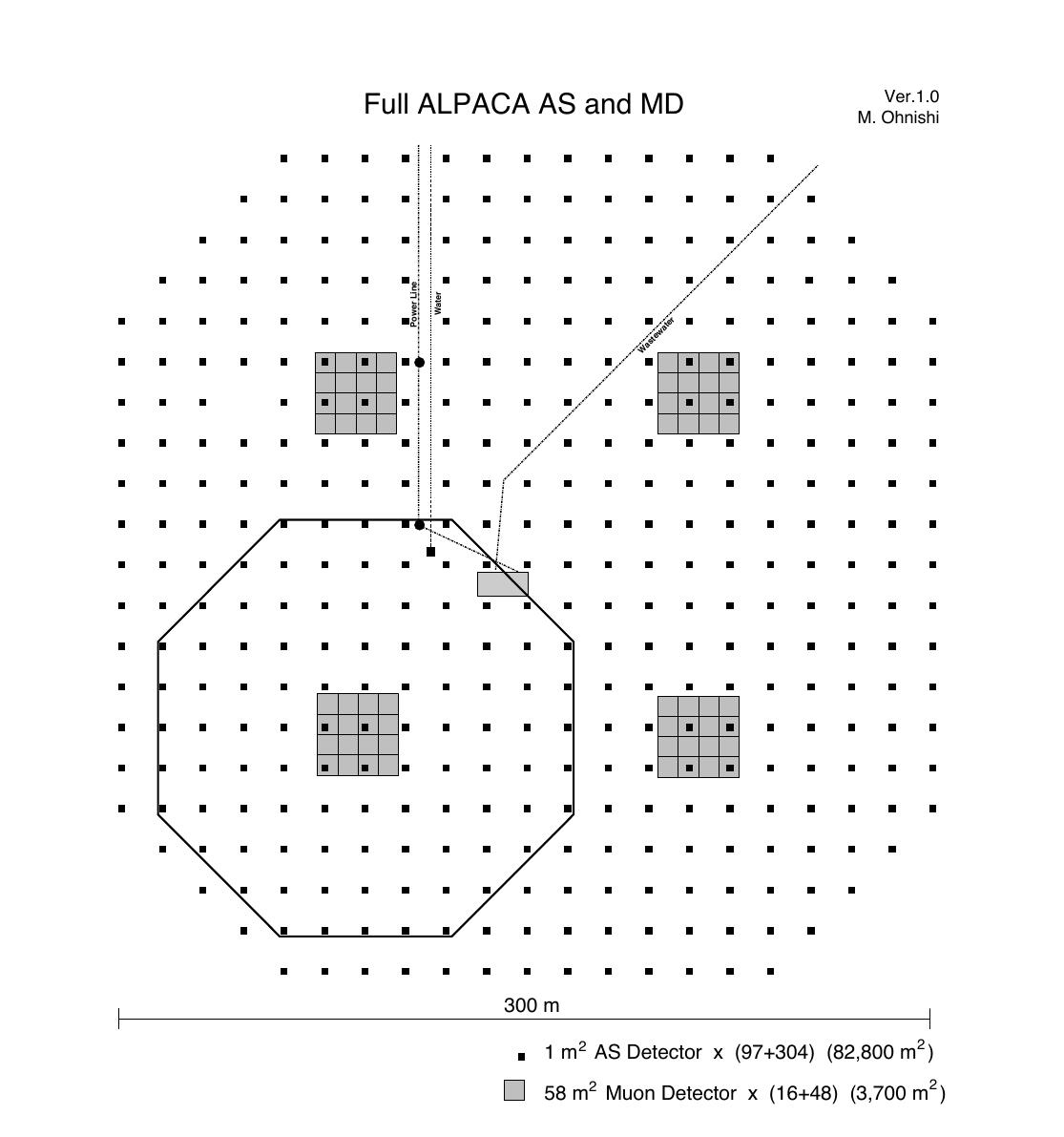}
\caption{Schematic view of the ALPACA array. The small black and large grey squares mean SDs and MDs, respectively. The gray rectangle in the center indicates the central electronics hut. The solid octagon in the lower left corner shows the location and area of the prototype array ALPAQUITA (see Section \ref{sec:alpaquita}).}
\label{view}
\end{figure}

ALPACA array consists of two main components; surface air shower detectors (SDs) and underground muon detectors (MDs) (see Figure \ref{view}). The array has a geometrical area of $\sim83,000\,\mathrm{m}^2$ and captures air showers produced by the interactions of primary gamma rays and cosmic rays with the atmosphere. Each SD consists of a $1\,\mathrm{m} \times 1\,\mathrm{m} \times 5\,\mathrm{cm}$ plastic scintillator and a 2-inch PMT covered by a steel box. 401\,SDs are installed at 15\,m intervals and can determine the energy and direction of arrival of a primary particle. On the other hand, MDs comprises water Cherenkov detectors constructed 2\,m underground and its total area is $\sim3,700\,\mathrm{m}^2$. Each MD consists of 16 cells, each of which have a area of $56\,\mathrm{m}^2$ and contains a water layer with a 1.5\,m depth. Most of the electromagnetic components in air showers are absorbed by the soil before reaching MDs, but only muons with energies above 1\,GeV penetrate into the water layer of the MDs and produce Cherenkov radiations, which are collected by a 20-inch PMT suspended downward at the cell’s ceiling after several reflection on the wall and floor. Counting the number of muons in air showers allow us to distinguish gamma-ray-induced air showers from (hadronic) cosmic-ray-induced ones. 

ALPACA's high duty cycle ($\sim100\,\%$) and wide field of view ($\sim2$ steradian) make it suitable for detecting sub-PeV gamma rays with low arrival frequencies. For 100 TeV gamma rays, the angular and energy resolutions are estimated at about 0.2\,degrees and 25\,\%, respectively. The hadron rejection capability of MDs is more than 99.9\,\% at 100 TeV, while keeping about 90\,\% of the gamma-ray efficiency.

\begin{figure}[h]
\centering
\includegraphics[width=0.67\textwidth]{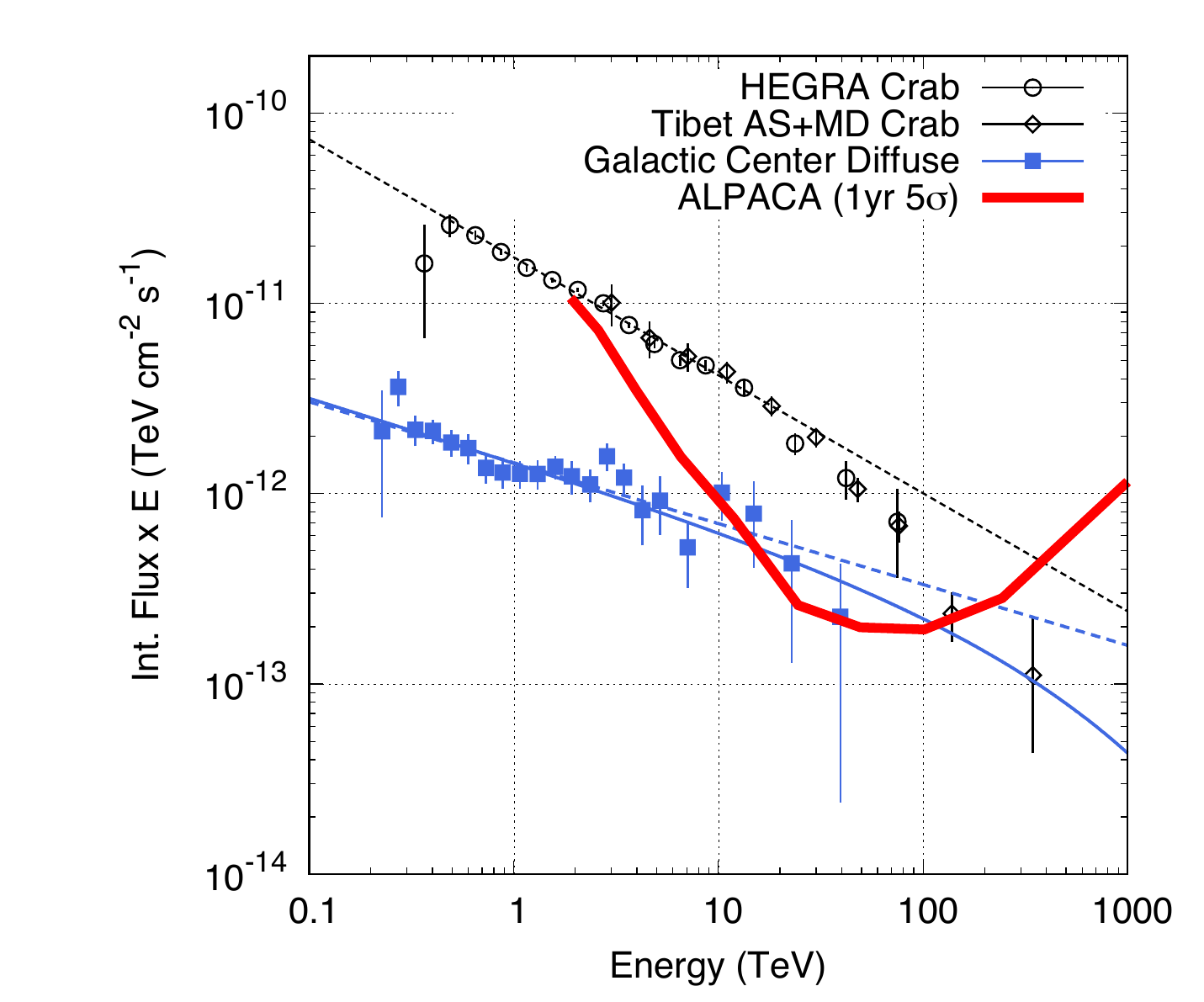}
\caption{Sensitivity curve of ALPACA (thick red curve) overlaid on the integral flux of the diffuse gamma rays from the Galactic Center region. The curve is estimated by MC simulation based on Tibet AS+MD within a zenith angle of 60\,degrees. Black points indicates the Crab Nebula spectrum measured by Tibet AS$\gamma$\cite{tibet_crab} and the dashed line is a power-law function with an exponential cutoff fitted to the spectrum.}
\label{alpaca_sense}
\end{figure}

Figure \ref{alpaca_sense} shows the sensitivity curve of ALPACA to gamma rays. The horizontal and vertical axis indicates the energies of gamma rays and the integral gamma-ray flux multiplied by energy. Blue points show the diffuse gamma-ray spectra in the Galactic Center region measured by H.E.S.S. telescope. The blue dashed and solid line indicates a power-law function (spectral index: -2.3) with an exponential cutoff and the hadronic emission model, respectively, fitted to the results of H.E.S.S., showing that ALPACA touches the Galactic Center flux above 100 TeV in one year.

\subsection{The prototype array: ALPAQUITA}
\label{sec:alpaquita}
\begin{figure}[H]
\centering
\includegraphics[width=0.61\textwidth]{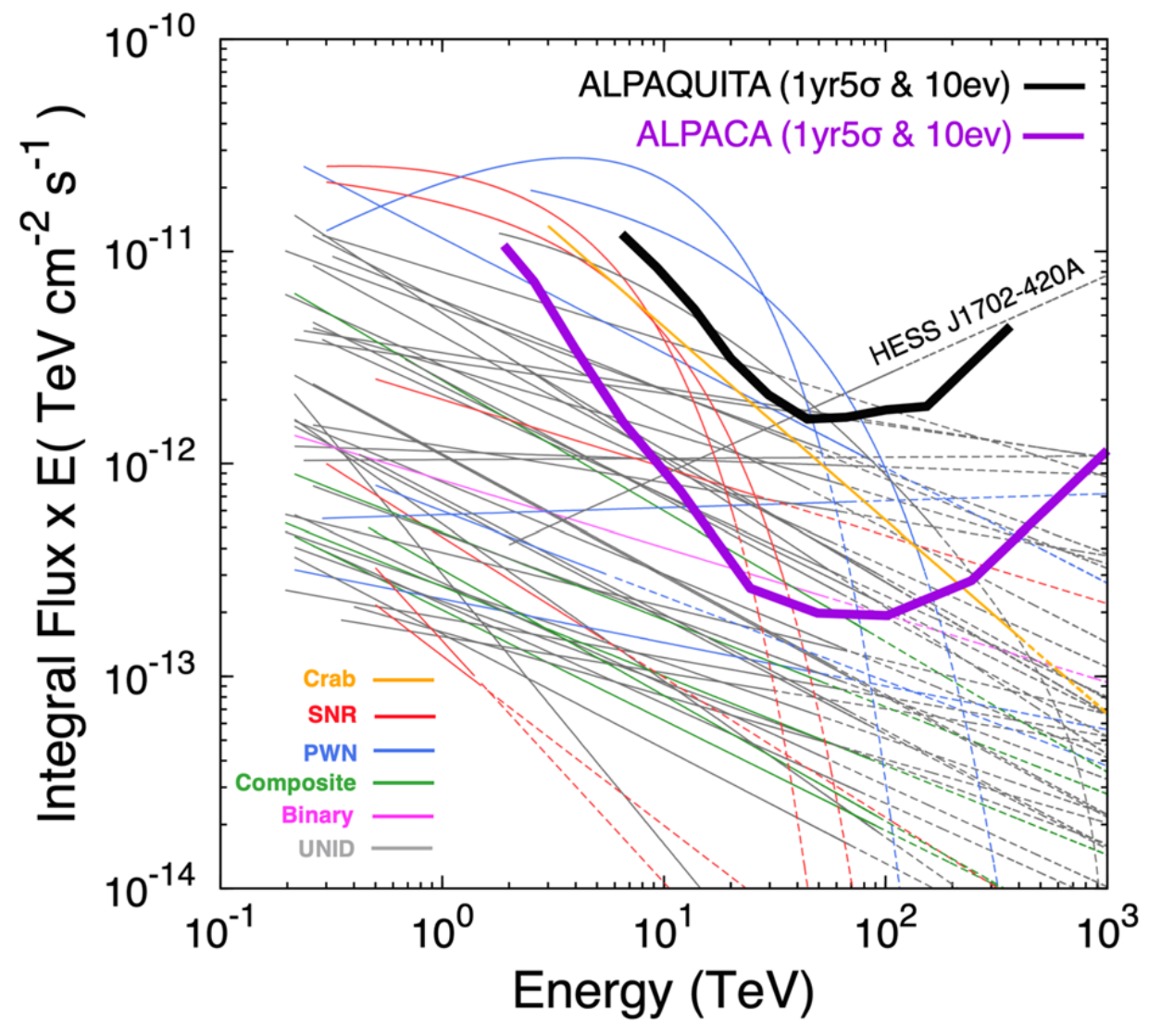}
\caption{Sensitivity curves of ALPAQUITA and ALPACA overlaid on the gamma-ray energy spectra of gamma-ray sources in the ALPACA field of view\cite{alpaquita_kato}. For each energy spectrum, the solid line represents the observed region by HESS\cite{hess_survey}\cite{hess_j1702} and HAWC\cite{hawc_multisources} and the dashed line shows the extrapolated region.}
\label{alpaquita_sense}
\end{figure}

The prototype-array ALPAQUITA consists of 97 SDs and one MD as shown in Figure \ref{view}. The ALPAQUITA construction is ongoing (see Section \ref{sec:construction}) and observations are scheduled to begin in 2022. Figure \ref{alpaquita_sense} shows the sensitivity curve of ALPAQUITA together with that of ALPACA sensitivity and the energy spectra of gamma-ray sources that resides in the ALPACA field of view. The thick black curve indicates that ALPAQTUIA can detect several gamma-ray sources including HESS J1702-420A in one year.

\section{Construction status of ALPAQUITA}
\label{sec:construction}
\begin{figure}[h]
\centering
\includegraphics[width=1.0\textwidth]{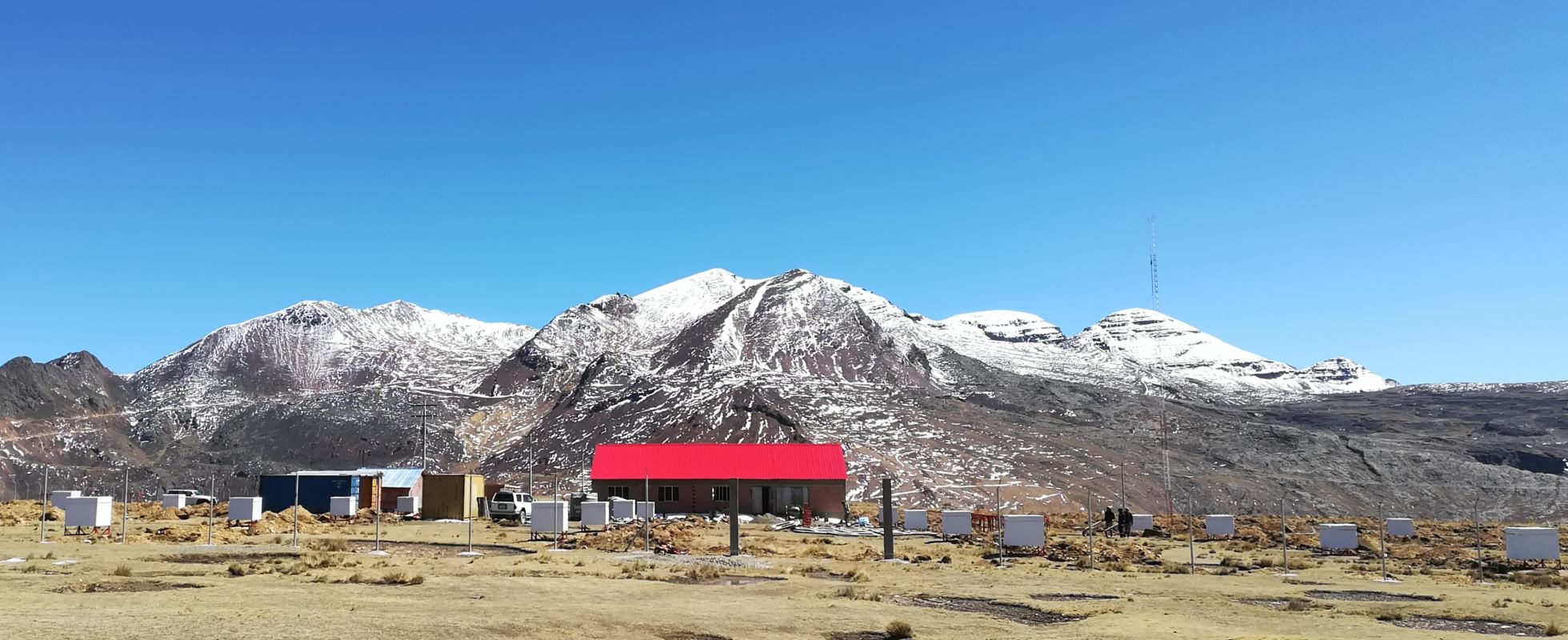}
\caption{Photo of ALPACA (ALPAQUITA) site. Mt. Chakartaya can be seen in the background. The building with the red roof is the central electronics hut. The multiple white cubes around the hut are SDs.}
\label{construction}
\end{figure}

The central electronics hut has already been built and 20\,SDs are already located around the hut (see Figure \ref{construction}). Installations of the perimeters, cables trenches, lightining rods, and internet Connection are also completed. ALPACA powerline is supplied by the branch of the Chacaltaya Observatory power line. We have restarted the construction of ALPAQUITA in June 2022 as the COVID-19 situation is improved.
\section{Conclusion}
Sub-PeV gamma-ray observaton is crutial to search PeV-cosmic-ray accelerators called PeVatrons. Tibet AS$\gamma$, HAWC, LHAASO Collaboration has pioneered sub-PeV energy gamma-ray astronomy in the Northern Hemisphere. The ALPACA experiment is a project aiming to search for sub-PeV gamma-ray sources in the Southern Hemisphere for the first time using a new air shower array in Bolivia. ALPACA is scheduled to start its observatio in 2024 and is expected to yield interesting results after one year of observations. ALPAQUITA, a prototype array under construction, can detect several bright sources in one-year observation such as HESS J1702-420A. The infrastructure for ALPAQUITA is being built locally and its DAQ is going to begin in 2022. Ultimately, the ALPACA project will identify Pevatrons in the southern sky and will help understand their acceleration mechanism of PeV comis rays.
\section*{Acknowledgements}
The ALPACA project is supported by the Japan Society for the Promotion
of Science (JSPS) through Grants-in-Aid for Scientific Research (A)
19H00678, Scientific Research (B) 19H01922, 
Scientific Research (B) 20H01920, 
Scientific Research (S) 20H05640, 
Scientific Research (B) 20H01234, 
Scientific Research (C) 22K03660, and 
Specially Promoted Research 22H04912, 
the LeoAtrox supercomputer located at the facilities of
the {\it Centro de An\'{a}lisis de Datos (CADS)}, CGSAIT, Universidad
de Guadalajara, M\'{e}xico, and by the joint research program of the
Institute for Cosmic Ray Research (ICRR), The University of Tokyo. 
K.~Kawata is supported by the Toray Science Foundation. 
E.~de la~Fuente thanks Coordinaci\'{o}n General Acad\'{e}mica y de
Innovaci\'{o}n (CGAI-UDG), cuerpo acad\'{e}mico PRODEP-UDG-CA-499,
Carlos Iv\'{a}n Moreno, Cynthia Ruano, Rosario Cedano, and Diana
Naylleli, for financial and administrative support during sabbatical
year stay at the ICRR on 2021. 
I.~Toledano-Juarez acknowledges support from CONACyT, M\'{e}xico; grant 754851.

\bibliography{SciPost_Example_BiBTeX_File.bib}

\nolinenumbers

\end{document}